\documentclass[twocolumn,showpacs,aps,prl,superscriptaddress]{revtex4}

\usepackage{graphicx}
\usepackage{dcolumn}
\usepackage{amsmath}
\usepackage{epsfig}

\RequirePackage{xspace}

\usepackage{relsize}
\def\babar{\mbox{\slshape B\kern-0.1em{\smaller A}\kern-0.1em
    B\kern-0.1em{\smaller A\kern-0.2em R}}}

\def\epem       {\ensuremath{e^+e^-}\xspace}

\def\Kbar  {\kern 0.2em\overline{\kern -0.2em K}{}\xspace}

\def\Kz    {\ensuremath{K^0}\xspace}
\def\Kzb   {\ensuremath{\Kbar^0}\xspace}
\def\KzKzb {\ensuremath{\Kz \kern -0.16em \Kzb}\xspace}
\def\Kp    {\ensuremath{K^+}\xspace}
\def\Km    {\ensuremath{K^-}\xspace}

\def\KpKm  {\ensuremath{\Kp \kern -0.16em \Km}\xspace}

\def\Dbar    {\kern 0.2em\overline{\kern -0.2em D}{}\xspace}

\def\Dz      {\ensuremath{D^0}\xspace}
\def\Dzb     {\ensuremath{\Dbar^0}\xspace}
\def\DzDzb   {\ensuremath{\Dz {\kern -0.16em \Dzb}}\xspace}
\def\Dp      {\ensuremath{D^+}\xspace}
\def\Dm      {\ensuremath{D^-}\xspace}

\def\DpDm    {\ensuremath{\Dp {\kern -0.16em \Dm}}\xspace}

\def\Bbar    {\kern 0.18em\overline{\kern -0.18em B}{}\xspace}

\def\BB      {\ensuremath{B\Bbar}\xspace} 
\def\Bz      {\ensuremath{B^0}\xspace}
\def\Bzb     {\ensuremath{\Bbar^0}\xspace}
\def\BzBzb   {\ensuremath{\Bz {\kern -0.16em \Bzb}}\xspace}
\def\Bu      {\ensuremath{B^+}\xspace}
\def\Bub     {\ensuremath{B^-}\xspace}

\def\BpBm    {\ensuremath{\Bu {\kern -0.16em \Bub}}\xspace}

\mathchardef\Upsilon="7107
\def\Y#1S{\ensuremath{\Upsilon{(#1S)}}\xspace}

\def\FourS {\Y4S}

\mathchardef\Deltares="7101
\mathchardef\Xi="7104
\mathchardef\Lambda="7103
\mathchardef\Sigma="7106
\mathchardef\Omega="710A

\def\Deltabar{\kern 0.25em\overline{\kern -0.25em \Deltares}{}\xspace}
\def\Lbar{\kern 0.2em\overline{\kern -0.2em\Lambda\kern 0.05em}\kern-0.05em{}\xspace}
\def\Sigbar{\kern 0.2em\overline{\kern -0.2em \Sigma}{}\xspace}
\def\Xibar{\kern 0.2em\overline{\kern -0.2em \Xi}{}\xspace}
\def\Obar{\kern 0.2em\overline{\kern -0.2em \Omega}{}\xspace}
\def\Nbar{\kern 0.2em\overline{\kern -0.2em N}{}\xspace}
\def\Xb{\kern 0.2em\overline{\kern -0.2em X}{}\xspace}

\def\BR         {{\ensuremath{\cal B}\xspace}}

\def\mes        {\mbox{$m_{\rm ES}$}\xspace}

\newcommand{\tev}{\ensuremath{\mathrm{\,Te\kern -0.1em V}}\xspace}
\newcommand{\gev}{\ensuremath{\mathrm{\,Ge\kern -0.1em V}}\xspace}
\newcommand{\mev}{\ensuremath{\mathrm{\,Me\kern -0.1em V}}\xspace}
\newcommand{\kev}{\ensuremath{\mathrm{\,ke\kern -0.1em V}}\xspace}
\newcommand{\ev}{\ensuremath{\mathrm{\,e\kern -0.1em V}}\xspace}
\newcommand{\gevc}{\ensuremath{{\mathrm{\,Ge\kern -0.1em V\!/}c}}\xspace}
\newcommand{\mevc}{\ensuremath{{\mathrm{\,Me\kern -0.1em V\!/}c}}\xspace}
\newcommand{\gevcc}{\ensuremath{{\mathrm{\,Ge\kern -0.1em V\!/}c^2}}\xspace}
\newcommand{\mevcc}{\ensuremath{{\mathrm{\,Me\kern -0.1em V\!/}c^2}}\xspace}

\def\invfb   {\ensuremath{\mbox{\,fb}^{-1}}\xspace}

\def\mus  {\ensuremath{\rm \,\mus}\xspace}

\def\mus        {\ensuremath{\,\mu{\rm s}}\xspace}    

\def\ra                 {\ensuremath{\rightarrow}\xspace}
\def\to                 {\ensuremath{\rightarrow}\xspace}

\newcommand{\stat}{\ensuremath{\mathrm{(stat)}}\xspace}
\newcommand{\syst}{\ensuremath{\mathrm{(syst)}}\xspace}

\def\pep2{PEP-II}

\def\gsim{{~\raise.15em\hbox{$>$}\kern-.85em
          \lower.35em\hbox{$\sim$}~}\xspace}
\def\lsim{{~\raise.15em\hbox{$<$}\kern-.85em
          \lower.35em\hbox{$\sim$}~}\xspace}

\def\CP                {\ensuremath{C\!P}\xspace}
\def\cp                {\ensuremath{C\!P}\xspace}

\xspace

\newcommand{\jprlBase}       {Phys.\ Rev.\ Lett.\xspace}
\newcommand{\jprBase}        {Phys.\ Rev.\xspace}
\newcommand{\jplBase}        {Phys.\ Lett.\xspace}

\newcommand{\nimBaseC}       {Nucl.\ Instr.\ and Methods\xspace}

\newcommand{\zpBase}         {Z.\ Phys.\xspace}

\newcommand{\nim}       [1]  {\nimBaseC~{\bf #1}}
\newcommand{\jplb}       [1]  {\jplBase\ B~{\bf #1}}

\newcommand{\jprl}      [1]  {\jprlBase\ {\bf #1}}
\newcommand{\jprd}      [1]  {\jprBase\ D~{\bf #1}}

\newcommand{\zp}        [1]  {\zpBase\ {\bf #1}}

\def\jetset74   {\mbox{\tt Jetset \hspace{-0.5em}7.\hspace{-0.2em}4}\xspace}

\def\figurebox#1#2#3{%
    \def\arg{#3}%
    \ifx\arg\empty
    {\hfill\vbox{\hsize#2\hrule\hbox to #2{\vrule\hfill\vbox to #1{\hsize#2\vfill}\vrule}\hrule}\hfill}%
    \else
    {\hfill\epsfbox{#3}\hfill}%
    \fi}

\newcommand{\btodsk}{$B^{-}\ra  D^{*0} K^{-}$}
\newcommand{\btodsp}{$B^{-}\ra  D^{*0} \pi^{-}$}

\newcommand{\btodsh}{$B^{-}\ra  D^{*0} h^{-}$}

\newcommand{\dotokp}{$D^0\ra K^-\pi^+$}
\newcommand{\dotokppp}{$D^0\ra K^-\pi^+\pi^+\pi^-$}
\newcommand{\dotokppo}{$D^0\ra K^-\pi^+\pi^0$}
\newcommand{\dotopp}{$D^0\ra \pi^-\pi^+$}
\newcommand{\dotokk}{$D^0\ra K^-K^+$}

\newcommand{\BBb}{$B\overline{B}$}
\newcommand{\pio}{$\pi^0$}
\newcommand{\deltae}{$\Delta E$}
\newcommand{\deltaemk}{$\Delta E_K$}

\newcommand{\Do}{D^0}

\begin{document}

\title{
{\large \bf \boldmath
Measurement of the Ratio \BR({\boldmath $B^{-}\ra  D^{*0}
K^{-}$})/\BR({\boldmath $B^{-}\ra  D^{*0} \pi^{-}$}) and of the \CP\ Asymmetry of
{\boldmath $B^-\ra D^{*0}_{\CP+}K^-$} Decays}
}

%
\author{B.~Aubert}
\author{R.~Barate}
\author{D.~Boutigny}
\author{F.~Couderc}
\author{Y.~Karyotakis}
\author{J.~P.~Lees}
\author{V.~Poireau}
\author{V.~Tisserand}
\author{A.~Zghiche}
\affiliation{Laboratoire de Physique des Particules, F-74941 Annecy-le-Vieux, France }
\author{E.~Grauges-Pous}
\affiliation{Universitad Autonoma de Barcelona, E-08193 Bellaterra, Barcelona, Spain }
% ...
\collaboration{The \babar\ Collaboration}
\noaffiliation

\date{\today}

\begin{abstract}
We study the decays \btodsp\ and \btodsk, where the
$D^{*0}$ decays into $D^0\pi^0$, with the $D^0$ reconstructed in the
\CP-even ($\CP+$) eigenstates $K^-K^+$ and $\pi^-\pi^+$ and in the (non-\CP) 
channels $K^-\pi^+$, $K^-\pi^+\pi^+\pi^-$, and $K^-\pi^+\pi^0$.
Using a
sample of about 123 million $B\overline{B}$ pairs, we measure the ratios of decay rates
\begin{displaymath}
R^{*}_{{\text {non-}}\CP}\equiv \frac{\BR(B^-\to D^{*0}_{{\text {non-}}\CP}K^-)}{\BR(B^-\to
D^{*0}_{{\text {non-}}\CP}\pi^-)} = 0.0813\pm0.0040\stat^{+0.0042}_{-0.0031}\syst,
\end{displaymath}
and provide the first measurements of
\begin{displaymath}
R^{*}_{\CP+}\equiv \frac{\BR(B^-\to D^{*0}_{\CP+}K^-)}{\BR(B^-\to D^{*0}_{\CP+}\pi^-)} = 0.086 \pm0.021 \stat \pm0.007 \syst,
\end{displaymath}
and of the \CP asymmetry
\begin{displaymath}
A^*_{\CP+}\equiv
\frac{\BR(B^-{\ra}D^{*0}_{\CP+}K^-)-\BR(B^+{\ra}D^{*0}_{\CP+}K^+)}{\BR(B^-{\ra}D^{*0}_{\CP+}K^-)+\BR(B^+{\ra}D^{*0}_{\CP+}K^+)}=
-0.10\pm0.23\stat^{+0.03}_{-0.04}\syst.
\end{displaymath}
\end{abstract}


\maketitle

The decays $B^-{\ra}D^{(*)0}K^{(*)-}$
will play an important role in our
understanding of \cp\ violation, as they can be used to constrain the
angle $\gamma=\arg(-V_{ud}V_{ub}^*/V_{cd}V_{cb}^*)$  of the 
Cabibbo-Kobayashi-Maskawa (CKM) matrix in a theoretically clean
way by exploiting the interference between the $b\ra c\overline{u}s$ and
$b\ra u\overline{c}s$ decay amplitudes~\cite{gronau1991}.
In the Standard Model, 
neglecting $D^{0}\overline{{D}^{0}}$ mixing,
$R^{*}_{\CP\pm}/R^{*}_{{\text {non-}}\CP}\simeq 1+r^2\pm2r\cos\delta
\cos\gamma$, where $\CP+(-)$ indicates \CP-even (odd) modes,
\begin{eqnarray}\label{eq:rstar}
R^{*}_{{\text {non-}}\CP/\CP\pm}\equiv \frac{\BR(B^-\ra
D^{*0}_{{\text {non-}}\CP/\CP\pm}K^-)}{\BR(B^-\ra D^{*0}_{{\text {non-}}\CP/\CP\pm}\pi^-)},
\end{eqnarray}
$r$ is the
absolute value of the ratio of the color suppressed $B^+\ra D^{*0}K^+$ and
color allowed $B^-\ra D^{*0}K^-$ amplitudes ($r \sim
0.1$--$0.3$), and $\delta$ is the strong phase difference between those
amplitudes. The decays \btodsp\ provide a convenient normalization term
since many systematic uncertainties are common to the two, while the
interference 
effects should be highly suppressed for the $D^{*0}\pi^-$, when
compared to the ones for the $D^{*0}K^-$ final states. Furthermore,
defining the direct \CP asymmetry 
\begin{eqnarray}\label{eq:cpa}
A^{*}_{\CP\pm}\equiv \frac{\BR(B^-{\ra}D^{*0}_{\CP\pm}K^-)-\BR(B^+{\ra}D^{*0}_{\CP\pm}K^+)}{\BR(B^-{\ra}D^{*0}_{\CP\pm}K^-)+\BR(B^+{\ra}D^{*0}_{\CP\pm}K^+)},
\end{eqnarray}
we have:
$A^*_{\CP\pm}=\pm 2r\sin\delta\sin\gamma/(1+r^2\pm
2r\cos\delta\cos\gamma)$. The unknowns $\delta$, $r$, and $\gamma$ can be
constrained by measuring $R^{*}_{{\text {non-}}\CP}$, $R^{*}_{\CP\pm}$, and
$A^{*}_{\CP\pm}$.
The Belle Collaboration has reported $R^{*}_{{\text {non-}}\CP} = 0.078\pm0.019\pm0.009$ using 10.1
\invfb\ of data~\cite{belle}.

We present the measurement of $R^*_{{\text {non-}}\CP}$, $R^*_{\CP+}$ and
$A^{*}_{\CP+}$ performed using 113 \invfb\ of data taken at the 
\FourS\ resonance by the \babar\ detector with the \pep2\ asymmetric $B$
factory. An additional 12 \invfb\
of data taken at a center-of-mass (CM) energy 40 MeV below the \FourS\ mass was
used for background studies.
The \babar\ detector is described in detail
elsewhere~\cite{detector}. 
Tracking of charged particles is provided by a five-layer silicon
vertex tracker (SVT) and a 40-layer drift chamber (DCH). The particle
identification exploits ionization energy loss in
the DCH and SVT, and Cherenkov photons detected in a ring-imaging
detector (DIRC). An electromagnetic
calorimeter (EMC), comprising 6580 thallium-doped CsI crystals,
is used to identify electrons and photons. 
These systems are mounted inside a 1.5-T solenoidal
superconducting magnet. Finally, the instrumented flux return (IFR) of the 
magnet allows discrimination of muons from other particles.
We use the GEANT4 Monte Carlo (MC)~\cite{geant4} program to simulate the
response 
of the detector, taking into account the varying
accelerator and detector conditions. 

We reconstruct \btodsh\
candidates, where the prompt 
track $h^-$ is a kaon 
or a pion. $D^{*0}$ candidates are reconstructed from 
$D^{*0}\ra D^0\pi^0$ decays and $D^0$ mesons from 
their decays to $K^-\pi^+$, $K^-\pi^+\pi^+\pi^-$, $K^-\pi^+\pi^0$,
$\pi^-\pi^+$, and $K^-K^+$. The first
three modes are referred to as ``non-\cp\ modes'', the
last two as ``\cp\ modes''.  Reference to the charge-conjugate decays is implied here and 
throughout the text, unless otherwise stated.

Charged tracks used in the reconstruction of $D$ and $B$ meson
candidates must have a
distance of closest approach to the 
interaction point less than $1.5$ cm in the transverse plane and less than 
$10$ cm along the beam axis. Charged tracks from the \dotopp\ decay
must also have transverse momenta greater than 0.1
\gevc and total momenta in the CM frame
greater than 0.25 \gevc. 
Kaon and pion candidates from all $D^0$
decays must pass particle identification (PID) selection
criteria, based on a neural-network algorithm which uses measurements of
${\rm d}E/{\rm d}x$ in the DCH and the SVT, and Cherenkov photons in the DIRC.

For the prompt track to be identified as a pion or a kaon,
we require that its Cherenkov angle  ($\theta_C$) be reconstructed 
with at least five photons. To suppress misreconstructed
tracks while maintaining high efficiency, events with prompt tracks
with $\theta_C$ more than 2 standard deviations (s.d.) away from the 
expected values for both the kaon and pion hypotheses are discarded; this
selection rejects most protons as well. The track is also discarded if it is
identified with high probability as an electron or a muon. 

Neutral pions are reconstructed by combining
pairs of photons with energy deposits larger
than 30 MeV in the calorimeter
that are not matched to charged tracks. The $\gamma\gamma$ invariant mass is
required 
to be in the range 122--146 \mevcc. The mass resolution for neutral
pions is typically 6--7 \mevcc. The minimum total laboratory energy
required for the $\gamma\gamma$ combinations 
is set to 200 MeV for \pio\ candidates from $D^0$ mesons. Only \pio\ candidates 
with CM momenta in the range 70--450 \mevc\ (denoted as soft
pions, $\pi_s$) are used to reconstruct 
the $D^{*0}$.

The $D^0$ mass resolution is 11 \mevcc for 
the \dotokppo\ mode and about 7 \mevcc\ for all other modes.
A mass-constrained fit is applied to the
$D$ candidate. 
The resolution of the difference
between the masses of the $D^{*0}$ and the daughter $D^{0}$ candidates
($\Delta M$) is typically in the range 0.8--1.0 \mevcc, depending on
the $\Do$ decay mode. 
A combined cut on the measured  
$D^0$ and soft-pion invariant masses and on $\Delta M$
is also applied by means of a $\chi^2$ defined as:
\begin{eqnarray}\label{eq:chi2}
\chi^2\!\equiv \!\left|\frac{m_{D^0}\!- \overline{m}_{D^0}
}{\sigma_{m_{D^0}}}\right|^2\!\!+\left|\frac{m_{\pi_s}\!- \overline{m}_{\pi_s}}
{\sigma_{m_{\pi_s}}}\right|^2 
\!\!+\left|\frac{\Delta M\!- \overline{\Delta M}}{\sigma_{\Delta M}}\right|^2\!\!,
\end{eqnarray}
where the mean values ($\overline{m}_{D^0}$, $\overline{m}_{\pi_s}$,
$\overline{\Delta M}$) and the resolutions ($\sigma_{m_{D^0}}$,
$\sigma_{m_{\pi_s}}$, $\sigma_{\Delta M}$) are measured 
in the data.
Correlations between the observables used in the $\chi^2$ in
Eq. (\ref{eq:chi2}) are negligible. Events with $\chi^2>9$ are rejected. 

$B$ meson candidates are reconstructed by combining a $D^{*0}$ candidate with a
high-momentum charged track. For the non-\CP\ modes, the charge of the
prompt track $h$ must match that of the kaon from the $D^0$ meson decay.
Two quantities are used to discriminate between signal and background: the
beam-energy-substituted mass  
$\mes \equiv \sqrt{(E_i^{*2}/2 + \mathbf{p}_i\cdot\mathbf{p}_B)^2/E_i^2-p_B^2}$
and the energy difference $\Delta E\equiv E^*_B-E_i^*/2$, 
where the subscripts $i$ and $B$ refer to the initial \epem\ system and the 
$B$ candidate respectively, and the asterisk denotes the CM frame. 

The \mes\ distribution for \btodsh\ signal can be described by
a Gaussian function centered at
the $B$ mass and does not depend on the nature of the prompt track. 
Its resolution, about $2.6$ \mevcc, is dominated by the uncertainty of
the beam energy and is slightly dependent on the $\Do$ decay mode.  
The observable \deltae\ does depend on the mass assigned to the
tracks forming the $B$ candidate, and on the \Dz\ momentum
resolution. 
We calculate \deltae\ with the kaon hypothesis for the prompt track and
indicate this quantity with \deltaemk.
For \btodsk\ events
\deltaemk\ is described approximately by a Gaussian centered at zero and with
resolution 17--18 \mev, whereas
for \btodsp\ events \deltaemk\ is shifted positively by about 50 \mev.
$B$ candidates with \mes\ in the range 5.2--5.3 \gevcc\ and with 
\deltaemk\ in the range ($-$100 to 130) \mev\ are selected.
 
A large fraction of the background consists of continuum (non
\BB) events and a 
powerful set of selection criteria is needed to suppress it. The selection 
is chosen to maximize the expected significance of the results,
based on MC studies. In the
CM frame, this background typically has 
two-jet structure, while \BB\ events are isotropic. 
We define $\theta_T$ as the angle between the thrust axes of the $B$
candidate and of the remaining charged and neutral 
particles in the event, both evaluated in the
CM frame, and signed so that the thrust axis component along
the  $e^-$ beam direction be positive. The distribution of
$|\cos\theta_T|$ is strongly peaked 
near one for continuum events and is 
approximately uniform for \BB\ events. For the non-\cp\ modes,
$|\cos\theta_T|$ is required to be less than $0.9$ for the \dotokp\ mode,
and less than $0.85$ for 
\dotokppp and \dotokppo\ modes for which the levels of the
continuum background are higher. For the \cp\ modes, $\cos\theta_T$
is required to be in the ranges ($-$0.9 to 0.85) and ($-$0.85 to 0.8) for the
\dotokk\ and \dotopp\ modes respectively. 
Other mode-dependent selection criteria are applied:  for the \dotokppp\ and
\dotokppo\ (\dotopp)
modes we reject events with 
$\cos\theta_{tD}<-0.9$ ($|\cos\theta_{tD}|>0.95$), where $\theta_{tD}$ is
the angle between the 
direction of the $\Do$\ in the laboratory and 
the opposite of the direction of the $K^-$ ($\pi^-$ for the \dotopp\ mode)
from the $\Do$\  
in the $\Do$\ rest frame. Finally, to reduce
combinatorial background in the \dotokppo\ final state, only those events
that fall in the enhanced regions of the Dalitz plots, according to the 
results of the Fermilab E691 experiment~\cite{dalitz2}, are selected. This
last requirement alone rejects 80\% of the background and accepts 69\% of the
signal, according to the MC simulation.

Multiple candidates are found in about 10--12\%
of the selected events with two- and four-body \Dz\ decays and in $17$\%
of the events with \dotokppo\ decays. 
The best candidate in each event is selected based on the  
$\chi^2$ previously defined. The number of candidates constructed with
the same $D^{*0}$, but different prompt track, is negligible; in this rare
case the best one in the event is randomly chosen.
The reconstruction efficiencies, based on MC simulation, are
reported in Table~\ref{tab:fitresults}. 

According to the simulation, the main contributions to the
\BB\ background for \btodsh\ events originate  
from the decays $B^-\rightarrow D^{(*)0}\rho^-$ and $B^0\rightarrow
D^{*-}h^+$. $B^-\rightarrow D^{*0}(\to D^0\gamma) h^-$ events are also
considered 
background as their \CP\ modes have \CP\ eigenvalues opposite to the
ones of the \btodsh signal~\cite{bondar}. 

For each \Dz\ decay
mode, an unbinned maximum-likelihood (ML) fit
is used to extract yields from the data for six candidate types: signal, 
continuum background,
and \BB\ background, for the kaon and pion choices for the mass
hypothesis of the prompt track in the candidate
decays \btodsh.

Three quantities from each selected candidate are used as input to the fit:
\deltaemk, \mes, and the $\theta_C$ of the prompt
track. The distributions of \deltaemk\ and \mes\ for the six 
candidate types are parametrized to build the probability density functions
(PDFs) that are used in the fit. 

Correlations between the \mes\ and \deltaemk\ variables for signal events
are about $-5$\% 
according to the simulation.
To account for these, we use signal MC events to parametrize the 
signal PDFs with a method  based on kernel
estimation~\cite{kernel}, which allows the description of a
two-dimensional PDF. 
The shapes of MC and data distributions of these observables are in 
good agreement, according to comparisons performed with pure samples of 
\btodsp\ events, obtained with very tight particle identification and 
kinematic selection. To the extent that we find differences in the data and 
MC distributions, we adjust the shapes of the PDFs to conform to the data. 
Systematic uncertainties due to limited statistics
associated with this procedure are included in the final results. 

We obtain the PDFs for the \mes\
distribution for continuum background from off-resonance
data, applying the standard selection criteria. The
\mes\ distributions are parametrized with a threshold
function~\cite{argus} defined as $f(\mes)\propto y
\sqrt{1-y^2}\exp{[-\xi(1-y^2)]}$,  
where $y=\mes/m_0$ and $m_0$ is the mean energy of the beams
in the CM frame.
The PDFs for the \deltaemk\ distributions for background
candidates from the continuum are 
well parametrized with exponential functions whose parameters are  
determined by
fitting the \deltaemk\  distributions of the selected \btodsh\ sample
in the off-resonance data. Both the \mes\ and the \deltaemk\ PDFs for the
continuum 
background are taken to be the same for \btodsp\ and \btodsk\ decays. The
shapes of MC and
data distributions of \mes\ and \deltaemk\ obtained with looser selection
criteria to increase the statistics, agree well for \btodsp\ and
\btodsk\ 
decays, validating this 
assumption. 
For the \cp\ modes very few off-resonance events pass the selection
criteria, hence we use the PDFs determined for the \dotokp\ mode.
This is justified by a separate comparison of the
\cp\ modes with the flavor-definite modes in data and MC samples obtained
with looser selection criteria.

The correlation between \mes\ and \deltaemk\ for the \BB\ background
is taken into account with a two-dimensional PDF determined
from simulated events, in a similar way to that used for the signal.

We obtain PDFs for the particle identification determination for the
prompt track from the distributions, in bins of momentum and polar angle,
of the difference between the reconstructed and expected 
$\theta_C$ of kaons and pions from \Dz\ decays in a control
sample that exploits the decay chain $D^{*+}\ra  
D^0\pi^+$, $D^0\ra K^-\pi^+$ to identify the tracks kinematically.

Initial PDFs are parametrized for each candidate type as detailed
above. With these we then fit pure samples of simulated signal events
and of background from off-resonance real and MC data. With the yields
from these fits we establish an efficiency matrix accounting for small
crossfeeds among the components. The corrections affecting the signal
yields are typically of order $1$\%.
The fractional systematic uncertainties for the signal yields associated
with these corrections are in the range 0.1--6.0\% depending on
the $\Do$ decay mode.

The likelihood $\mathcal{L}$ for the selected sample is given by the
product of the final PDFs for each individual candidate and a Poisson
factor:
\begin{eqnarray}\label{eq:pdf}
\mathcal{L}\equiv \frac{e^{-N'}(N')^N}{N!}\prod_{i=1}^{N}\sum_{j=1}^6
\frac{N_{j}}{N'}\mathcal{P}_j(m_{{\rm {ES}}_i},\Delta E_{K_i},\theta_{C_i}) 
\end{eqnarray}
where $N$ is the total number of events, $N_{j}$ are the yields for each
of the previously defined six candidate types, and $N'\equiv \sum_{j=1}^6 N_{j}$,
$\mathcal{P}_j(m_{{\rm {ES}}_i},\Delta E_{K_i},\theta_{C_i})$ is the 
probability to measure the particular set of 
physical quantities ($m_{{\rm {ES}}_i}$,$\Delta E_{K_i}$,$\theta_{C_i}$) in the
$i^{th}$ event for a 
candidate of type $j$. The Poisson factor is the probability of observing
$N$ total events when $N'$ are expected.
The quantity $\mathcal{L}$ is maximized with respect to the
six yields using
the MINUIT program~\cite{minuit}. The fit has also been performed on
luminosity-weighted MC  
and high statistics toy MC events and it has been found to be unbiased.

The results of the fit are reported in detail in Table~\ref{tab:fitresults}. 
These yields are used to determine the \CP\ asymmetry parameters.
We measure:
\begin{eqnarray}
R^*_{{\text {non-}}\CP}       &=& 0.0813\pm0.0040\stat^{+0.0042}_{-0.0031}\syst, \nonumber\\
R^*_{\CP+} &=& 0.086 \pm0.021 \stat \pm0.007 \syst, \nonumber\\
R^*_{\CP+}/R^*_{{\text {non-}}\CP} &=& 1.06\pm0.26\stat^{+0.10}_{-0.09}    \syst, \nonumber\\
A^*_{\CP+}    &=& -0.10\pm0.23\stat^{+0.03}_{-0.04}\syst. \nonumber
\end{eqnarray}

\begin{figure}[!htb]
\begin{center}
\includegraphics[width=8.7cm]{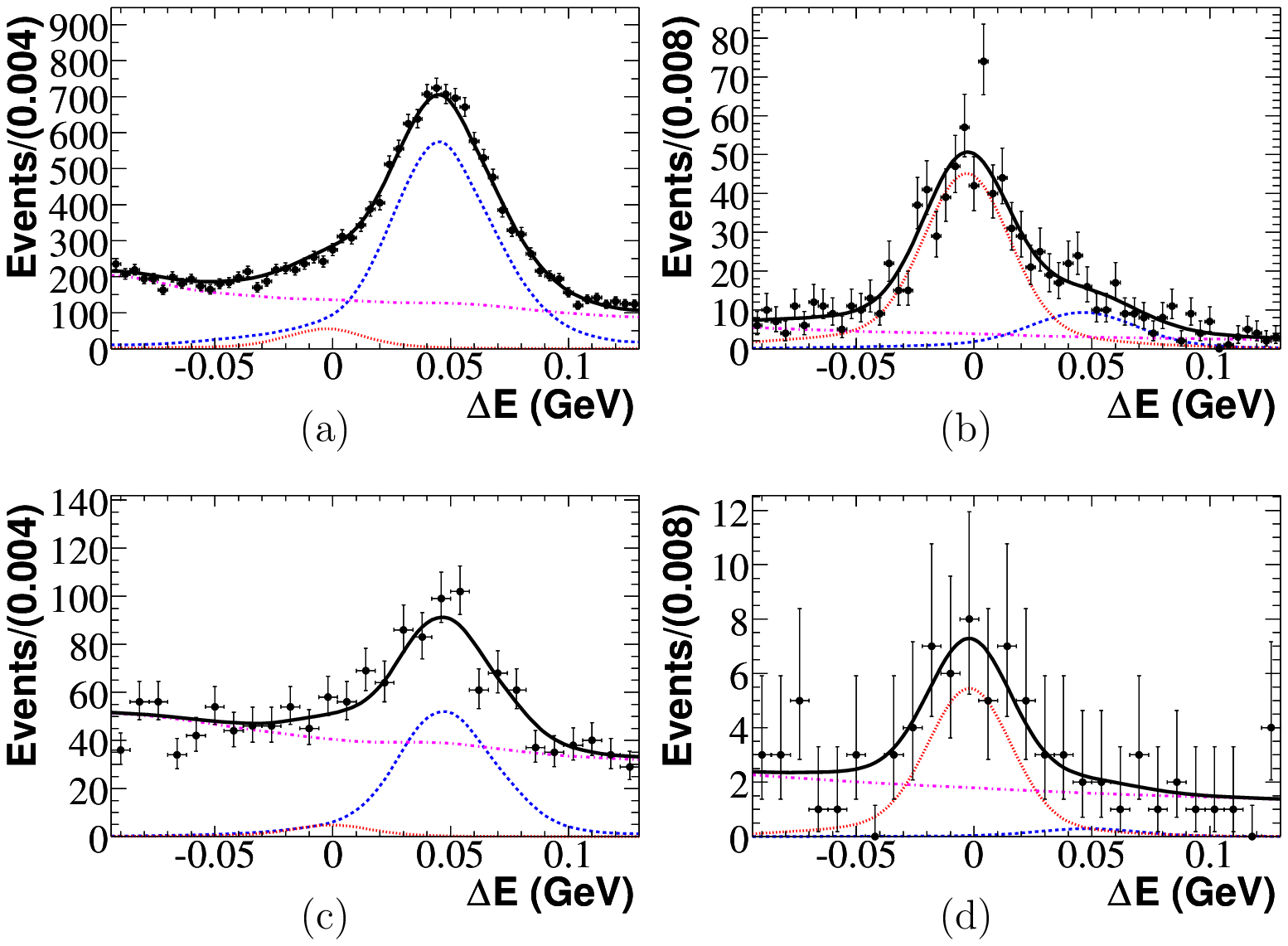}
\caption{Distributions of \deltaemk\ in the $B\rightarrow D^{*0}h$ sample, for
$D^0\rightarrow K^-\pi^+, K^-\pi^+\pi^0, K^-\pi^+\pi^+\pi^-$ ((a),
(b)) and  $D^0\rightarrow K^-K^+, \pi^-\pi^+$ ((c), (d)), before
((a), (c)) and after ((b), (d)) enhancing the $B\rightarrow
D^{*0}K$ component by requiring that the prompt track be consistent with the
kaon hypothesis and $\mes>5.27$\gevcc.  The
\btodsp\ signal contribution on the right of each plot is shown as a
dashed line,  
the \btodsk\ signal on the left as a dotted line, and the background
as a dashed-dotted line. The total fit with all the contributions is
shown with a thick solid line.}
\label{fig:fit}
\end{center}
\end{figure}

Figure~\ref{fig:fit} shows the distributions of \deltaemk\ for the 
combined non-\CP\ and \CP\ modes before and after the enhancement of the
$B\rightarrow 
D^{*0}K$ component. The enhancement is accomplished by requiring that the
prompt track be consistent with the 
kaon hypothesis and that $\mes>5.27$\gevcc. The \deltaemk\
projections of the fit results are also shown.
 
\begin{table}[ht]
\caption{Results of the yields from the ML fit. For the
\cp\ modes the results of the fit separately for the $B^+$ and $B^-$
samples are also quoted. Errors are statistical only. The efficiencies
($\epsilon$) based on MC simulation are
also reported.}
\label{tab:fitresults}
\begin{center}
\begin{tabular}{|l|c|c|c|}
\hline
&&&\\[-9pt]
$D^0$ mode &  $N(B\rightarrow D^{*0}\pi)$ & $N(B\rightarrow D^{*0}K)$ &
$\varepsilon(D^{*0}\pi$) (\%)\\
\hline
\hline
$K^-\pi^+$           &\  $2639\pm 56$ &\  $226  \pm 18$  &\ $17.5\pm0.2$\\
$K^-\pi^+\pi^0$      &\  $3249\pm 68$ &\  $247  \pm 21$  &\ $ 5.9\pm0.1$\\
$K^-\pi^+\pi^+\pi^-$ &\  $3071\pm 64$ &\  $242  \pm 21$  &\ $ 9.7\pm0.1$\\ 
\hline 
\hline
$K^-K^+$             &\  $258 \pm 19$ &\  $23.4 \pm 5.6$ &\ $15.3\pm0.2$\\
$K^-K^+$ [$B^+$]     &\  $123 \pm 13$ &\  $13.4 \pm 4.1$ &\ $15.6\pm0.3$\\
$K^-K^+$ [$B^-$]     &\  $134 \pm 13$ &\  $ 9.9 \pm 3.7$ &\ $14.9\pm0.3$\\
\hline
$\pi^-\pi^+$         &\  $124 \pm 14$ &\  $ 6.3 \pm 4.6$ &\ $14.6\pm0.2$\\
$\pi^-\pi^+$ [$B^+$] &\  $75  \pm 11$ &\  $ 0.7 \pm 3.2$ &\ $14.5\pm0.3$\\
$\pi^-\pi^+$ [$B^-$] &\  $49  \pm  9$ &\  $ 5.3 \pm 3.5$ &\ $14.8\pm0.3$\\
\hline
\end{tabular}
\end{center}
\end{table}

The ratio of the decay rates for \btodsp\ and \btodsk\ is
separately calculated for the different $D^0$ decay channels and is
computed with the signal yields estimated with the
ML fit and listed in 
Table~\ref{tab:fitresults}. The resulting ratios are scaled by
correction factors of a few percent, which are estimated with simulated data
and which take into account small differences in the efficiency between
\btodsk\ and \btodsp\ event selections. The results are listed in
Table~\ref{tab:final_ratio}. 

\begin{table}[ht]
\caption{Measured ratios
for different \Dz\ decay modes. The first error is statistical, the second is
systematic.} 
\label{tab:final_ratio}
\begin{center}
\begin{tabular}[t]{|l|c|}
\hline
&\\[-9pt]
\btodsh\ Mode &\ \textbf{\BR($B{\ra}D^{*0}K$)/\BR($B{\ra}D^{*0}\pi$)} $(\%)$\\
\hline
\hline
&\\[-9pt]
\dotokp&\ \ \ \ \ \        $8.93\pm 0.72^{+0.38}_{-0.30}$\\
\dotokppo&\ \ \ \ \ \      $7.59\pm 0.65^{+0.37}_{-0.27}$\\
\dotokppp&\ \ \ \ \ \      $7.91\pm 0.72^{+0.61}_{-0.59}$\\
\hline
&\\[-9pt]
Weighted Mean (non-\CP)&\ \ \ \ \ \  $8.13\pm 0.40^{+0.42}_{-0.31}$\\
\hline
\hline
&\\[-9pt]
\dotokk &\ \ \ \ \ \       $9.4\pm 2.3\pm0.6$\\
\dotopp &\ \ \ \ \ \       $5.9\pm 4.4^{+1.0}_{-1.4}$\\
\hline
Weighted Mean (\CP)&\ \ \ \ \ \  $8.6\pm 2.1\pm0.7$\\
\hline
\end{tabular}
\end{center}
\end{table}

\begin{table}[ht]
\caption{Average systematic uncertainties for $R^*_{({\text {non-}})\CP}$.} 
\begin{center}
\begin{tabular}{|l|c|c|}
\hline
Systematic & $\Delta R^*_{{\text {non-}}\CP}/R^*_{{\text {non-}}\CP} (\%)$& $\Delta R^*_{\CP}/R^*_{\CP} (\%)$\\
Source   &  non-\CP\ modes &  \CP\ modes   \\
\hline
\hline
&&\\[-9pt]
$\Delta E_K $ (signal)       &$^{+  2.0}_{-  1.8}$   &$^{+  2.7}_{-  2.8}$  \\
$\Delta E_K(q\bar q)$        &$^{+  0.3}_{-  0.6}$   &$^{+  0.9}_{-  2.5}$  \\
$\Delta E_K(B\bar B)$        &$^{+  0.0}_{-  0.5}$   &$^{+  0.8}_{-  0.8}$  \\
\mes (signal)                &$^{+  0.4}_{-  0.3}$   &$^{+  0.7}_{-  0.4}$  \\
\mes $ (q\bar q)$            &$^{+  0.8}_{-  0.8}$   &$^{+  4.4}_{-  6.7}$  \\
\mes $ (B\bar B)$            &$^{+  1.2}_{-  1.3}$   &$^{+  0.3}_{-  3.2}$  \\
PDF Crossfeeds	             &$^{+  2.8}_{-  2.6}$   &$^{+  0.7}_{-  0.7}$  \\
PID PDF                      &$^{+  3.0}_{-  1.8}$   &$^{+  4.0}_{-  1.4}$  \\
$\varepsilon$ Correction     &$^{+  1.4}_{-  1.4}$   &$^{+  2.0}_{-  2.0}$  \\
\hline
\end{tabular}
\end{center}
\label{tab:syst_ratio}
\end{table}

The sources of systematic uncertainties for the yields have been identified and their
contributions (for the measurement of $R^*_{({\text {non-}})\CP}$) are reported in
Table~\ref{tab:syst_ratio}. 
Uncertainties of the signal parametrizations of \deltaemk\ and \mes\ arise
from the assumed shapes of the
PDFs and discrepancies between real and simulated data.  
All the parameters of the \deltaemk\ and \mes\ PDFs have also been varied
according to their one s.d. statistical uncertainties and signed 
variations in the yields are taken as systematic
uncertainties. 
For the \BBb\ and continuum backgrounds, the
systematic uncertainties due to the limited statistics of the MC and of the
off-resonance data  
have been calculated varying the \deltaemk\ and \mes\ PDF parameters by their
statistical uncertainties.
There are several contributions to the PID systematic uncertainty for the 
prompt track: the uncertainty
due to limited statistics is calculated by varying each
parameter of the PDF, in each bin in momentum and polar angle, by its
uncertainty (keeping constant all other parameters in the same bin 
and all parameters in all the other bins) and summing all the contributions 
in quadrature; results obtained with alternative PID PDFs, which account
for different 
$\theta_C$ residual
shapes and for discrepancies between data and simulation, are also included as
systematic 
uncertainties. The systematic uncertainties due to the fit crossfeeds  
have been evaluated. Finally, errors associated with the efficiency correction factor
are also included. 

Many of the systematic uncertainties for the signal yields have similar
effects  
on the \btodsk\ and \btodsp\ events (they increase or decrease both fractions
simultaneously), hence their effect is reduced in deriving the
systematic uncertainty for the measurement of the ratios, when all
correlations are taken into 
account. 
Overall, the main sources of systematic
uncertainties for the
measurement of both $R^*_{({\text {non-}})\CP}$ and $A^*_{\CP+}$ are due to the characterization
of the shapes of \mes\ and \deltaemk\ for the signal, to the
characterization of the \mes\ PDFs for the background, to the particle
identification, and to the uncertainty of the fit crossfeeds and of the efficiency
correction factors. The systematic uncertainty for $A^*_{\CP+}$ due to
possible detector charge asymmetries is evaluated by measuring asymmetries
analogous 
to those defined in Eq. (\ref{eq:cpa}), but for \btodsp\ and \btodsk\ events (the latter
uniquely for the non-\CP\ modes), where \CP\ violation is
expected to be negligible. Results for all modes are then combined, taking
correlations into account. The measured asymmetry is
$-0.008\pm0.012\stat\pm0.001\syst$. Though it is consistent with zero, it
is also consistent with $-0.020$ at one s.d. level, hence we take the
magnitude of this value
as a further symmetric systematic uncertainty on $A^*_{\CP+}$.
When combining the  
results for the different modes, all systematic
and statistical uncertainties are considered to be uncorrelated, except
for the contributions of the
PID PDF (common to all modes) and of the detector charge asymmetry in the
measurement of $A^*_{\CP+}$, which are 
considered to be completely correlated. For the measurement of $R^*_{\CP+}/R^*_{{\text {non-}}\CP}$
all systematic uncertainties have been considered to be uncorrelated; this
assumption is conservative, and has negligible effect 
on the final result, which is largely statistically limited.

In conclusion, we have measured the ratio of the decay rates for
$B^-\ra D^{*0}K^-$ and $B^-\ra D^{*0}\pi^-$ processes, with
non-\CP\ eigenstates. This constitutes the most
precise measurement for this channel. We have also performed the first
measurement of 
the same ratio and of the \cp\
asymmetry $A^*_{\CP+}$ for $D^0$ mesons decaying to \cp\ eigenstates. These
results, together with measurements exploiting $B^-\ra D^{0}K^-$, $B^-\ra
D^{0}K^{*-}$ and $B^-\ra
D^{*0}K^{*-}$ decays~\cite{belle,expres}, constitute a first step
towards measuring the 
angle $\gamma$. Furthermore, assuming factorization and 
flavor-SU(3) symmetry, theoretical calculations (in the tree-level
approximation) predict: $\BR(B^-\ra D^{*0}K^-)/\BR(B^-\ra
D^{*0}\pi^-)\sim(V_{us}/V_{ud})^2(f_K/f_\pi)^2\sim$0.074, where $f_K$
and$f_\pi$ are the meson decay constants~\cite{RinMS}. Our results accord
with these predictions.

We are grateful for the excellent luminosity and machine conditions
provided by our \pep2\ colleagues, 
and for the substantial dedicated effort from
the computing organizations that support \babar.
The collaborating institutions wish to thank 
SLAC for its support and kind hospitality. 
This work is supported by
DOE
and NSF (USA),
NSERC (Canada),
IHEP (China),
CEA and
CNRS-IN2P3
(France),
BMBF and DFG
(Germany),
INFN (Italy),
FOM (The Netherlands),
NFR (Norway),
MIST (Russia), and
PPARC (United Kingdom). 
Individuals have received support from CONACyT (Mexico), A.~P.~Sloan Foundation, 
Research Corporation,
and Alexander von Humboldt Foundation.

%
%


\begin{thebibliography}{99}

\bibitem{gronau1991} M.~Gronau and D.~Wyler, \jplb{265}, 172 (1991);
  M.~Gronau and D.~London, \jplb{253}, 483 (1991); D.~Atwood, I.~Dunietz,
  and A.~Soni, \jprl{78}, 3257 (1997); A.~Soffer, \jprd{60}, 054032 (1999);
  M.~Gronau, \jprd{58}, 037301 (1998); Z.~Xing, \jprd{58}, 093005 (1998);
  J.H.~Jang and P.~Ko, \jprd{58}, 111302 (1998); M.~Gronau and
  J.L.~Rosner, \jplb{439}, 171 (1998).
\bibitem{belle} Belle Collaboration, K.~Abe $et\ al.$, \jprl{87}, 111801 (2001).
\bibitem{detector} \babar\ Collaboration, B.~Aubert {\em et al.}, \nim{A479}, 1 (2002).
\bibitem{geant4} GEANT4 Collaboration, S.~Agostinelli {\em et al.}, \nim{A506}, 250 (2003). 
\bibitem{dalitz2} E691 Collaboration, J.~C.~Anjos {\em et al.}, \jprd{48}, 56 (1993).
\bibitem{bondar} A.~Bondar and T.~Gershon, \jprd{70}, 091503 (2004).
\bibitem{kernel} K.~S.~Cranmer, Comp. Phys. Commun. {\bf 136}, 198 (2001).
\bibitem{argus} ARGUS Collaboration, H.~Albrecht {\em et al.}, \zp{C48}, 543 (1990).
\bibitem{minuit} F.~James, Comput. Phys. Commun. {\bf 10}, 343 (1975).
\bibitem{expres} \babar\ Collaboration, B.~Aubert {\em et al.}, \jprd{69}, 051101 (2004);
\babar\ Collaboration, B.~Aubert {\em et al.}, \jprl{92}, 202002 (2004); \babar\ Collaboration, B.~Aubert {\em et al.}, \jprl{92} 141801 (2004);
Belle Collaboration, S.K.~Swain {\em et al.}, \jprd{68}, 051101 (2003).
\bibitem{RinMS} M.~Gronau {\em et al.}, \jprd{52}, 6356 (1995).

\end{thebibliography}
\end{document}